\documentclass[11pt,letterpaper]{extarticle}

\usepackage{typearea}
\typearea{15} 

\usepackage{shadow}
\usepackage{theorem,latexsym,graphicx,amssymb}
\usepackage{amsmath,enumerate}
\usepackage{wrapfig}
\usepackage{epsfig}
\usepackage{xspace}
\usepackage{bm}
\usepackage{layout}
\usepackage{shadethm}
\usepackage[compact]{titlesec}
\usepackage{algorithm, algorithmic}
\usepackage{color}
\usepackage{ifpdf}
\usepackage[square,numbers]{natbib}

\usepackage{booktabs}
\usepackage{color}
\usepackage{colortbl}

\usepackage{float}

\newcommand{\denselist}{\itemsep 0pt\topsep-6pt\partopsep-6pt}

\definecolor{Darkblue}{rgb}{0,0,0.4}
\definecolor{Brown}{cmyk}{0,0.81,1.,0.60}
\definecolor{Purple}{cmyk}{0.45,0.86,0,0}
\newcommand{\mydriver}{hypertex}
\ifpdf
 \renewcommand{\mydriver}{pdftex}
\fi
\usepackage[breaklinks,\mydriver]{hyperref}
\hypersetup{colorlinks=true,
            citebordercolor={.6 .6 .6},linkbordercolor={.6 .6 .6},%
citecolor=Darkblue,urlcolor=black,linkcolor=Darkblue,pagecolor=black}
\newcommand{\lref}[2][]{\hyperref[#2]{#1~\ref*{#2}}}


\floatstyle{ruled}
\newfloat{algo}{tbp}{lop}[section]
\floatname{algo}{Algorithm}

\newcommand{\Z}[0]{{\ensuremath{\mathbb{Z}}}}

\newcommand{\vareps}[0]{\varepsilon}

\newcommand{\opt}{\ensuremath{{\sf opt}}\xspace}

\newtheorem{theorem}{Theorem}[section]

\newtheorem{lemma}[theorem]{Lemma}
\newtheorem{fact}[theorem]{Fact}
\newtheorem{claim}[theorem]{Claim}

\theoremstyle{plain}
\newtheorem{definition}[theorem]{Definition}
\theoremstyle{plain}

\newcommand{\RNC}{{\sf RNC}\xspace}
\newcommand{\NC}{{\sf NC}\xspace}
\newcommand{\prob}[1]{\ensuremath{\text{{\bf Pr}$\left[#1\right]$}}}
\newcommand{\expct}[1]{\ensuremath{\text{{\bf E}$\left[#1\right]$}}}

\newcommand{\kmed}{\mbox{\sc kMed}\xspace}
\newcommand{\kcenter}{\mbox{\sc kCenter}\xspace}
\newcommand{\kmeans}{\mbox{\sc kMeans}\xspace}
\newcommand{\facloc}{\mbox{\sc FacLoc}\xspace}
\newcommand{\fstar}{\mathcal{S}}
\newcommand{\price}{{\sf price}\xspace}
\newcommand{\udom}{\mbox{\sc MaxUDom}\xspace}
\newcommand{\maxdom}{\mbox{\sc MaxDom}\xspace}

\newenvironment{proof}{\noindent{\em Proof:} }{\hfill$\blacksquare$}

\usepackage{microtype}

\begin{document}
\title{Parallel Approximation Algorithms for Facility-Location
  Problems\thanks{Computer Science Department, Carnegie Mellon
    University, Pittsburgh, PA 15213.}}  

\author{{Guy E. Blelloch \qquad\qquad Kanat Tangwongsan}}

\date{}

\begin{titlepage}
  \def\thepage{}
  \thispagestyle{empty}
  
  \maketitle
  
  \begin{abstract}
    This paper presents the design and analysis of parallel
    approximation algorithms for facility-location problems, including
    $\NC$ and $\RNC$ algorithms for (metric) facility location,
    $k$-center, $k$-median, and $k$-means.  These problems have
    received considerable attention during the past decades from the
    approximation algorithms community, concentrating primarily on
    improving the approximation guarantees. In this paper, we ask,
    \emph{is it possible to parallelize some of the beautiful results
      from the sequential setting?}

    \smallskip Our starting point is a small, but diverse, subset of
    results in approximation algorithms for facility-location
    problems, with a primary goal of developing techniques for
    devising their efficient parallel counterparts.  We focus on
    giving algorithms with low depth, near work efficiency (compared
    to the sequential versions), and low cache complexity. Common in
    algorithms we present is the idea that instead of picking only the
    most cost-effective element, we make room for parallelism by
    allowing a small slack (e.g., a $(1+\vareps)$ factor) in what can
    be selected---then, we use a clean-up step to ensure that the
    behavior does not deviate too much from the sequential steps.  In
    this paper, we first present a parallel \RNC algorithm mimicking
    the greedy algorithm of Jain et al.~(\textit{J. ACM,
      50(6):795--824, 2003.}).  This is the most challenging algorithm
    to parallelize because the greedy algorithm is inherently
    sequential. We show the algorithm gives a
    $(3.722+\varepsilon)$-approximation and does $O(m
    \log_{1+\varepsilon}^2 m)$ work, which is within a logarithmic
    factor of the serial algorithm. Then, we present a simple $\RNC$
    algorithm using the primal-dual approach of Jain and
    Vazirani~(\textit{J. ACM, 48(2):274--296, 2001.}), which leads to
    a $(3+\varepsilon)$-approximation, and for input of size $m$ runs
    in $O(m \log_{1+\varepsilon} m)$ work, which is the same as the
    sequential work.  The sequential algorithm is a $3$-approximation.
    Following that, we present a local-search algorithm for $k$-median
    and $k$-means, with approximation factors of $5 + \vareps$ and $81
    + \vareps$, matching the guarantees of the sequential
    algorithms. For constant $k$, the algorithm does $O(n^2\log n)$
    work, which is the same as the sequential
    counterpart. Furthermore, we present a $2$-approximation algorithm
    for $k$-center with $O((n\log n)^2)$ work, based on the algorithm
    of Hochbaum and Shmoys~(\textit{Math. OR, 10(2):180--184, 1985.}).
    Finally, we show a $O(m\log^2_{1+\vareps}(m))$-work randomized
    rounding algorithm, which yields a $(4+\vareps)$-approximation,
    given an optimal linear-program solution as input.  The last two
    algorithms run in work within a logarithmic factor of the serial
    algorithm counterparts.  All these algorithms are ``cache
    efficient'' in that the cache complexity is bounded by $O(w/B)$,
    where $w$ is the work in the \textsf{EREW} model and $B$ is the
    block size.
  \end{abstract}
\end{titlepage}
\section{Introduction}
\label{sec:intro}

Facility location is an important and well-studied class of problems
in approximation algorithms, with far-reaching implications in areas
as diverse as machine learning, operations research, and networking:
the popular $k$-means clustering and many network-design problems are
all examples of problems in this class.  Not only are these problems
important because of their practical value, but they appeal to study
because of their special stature as ``testbeds'' for techniques in
approximation algorithms. Recent research has focused primarily on
improving the approximation guarantee, producing a series of beautiful
results, some of which are highly efficient---often, with the
sequential running time within constant or polylogarithmic factors of
the input size.

Despite significant progress on these fronts, work on developing
parallel approximation algorithms for these problems remains virtually
non-existent.  Although variants of these problems have been
considered in the the distributed computing
setting~\cite{MoscibrodaW:podc05,GehweilerLS:SPAA06,PanditP:podc09},
to the best our of knowledge, almost no prior work has looked directly
in the parallel setting where the total work and parallel time (depth)
are the parameters of concern. The only prior work on these problems
is due to Wang and Cheng, who gave a $2$-approximation algorithm for
$k$-center that runs in $O(n \log^2 n)$ depth and $O(n^3)$
work~\cite{WangC:PDP90}, a result which we improve in this paper. 


Deriving parallel algorithms for facility location problems is a
non-trivial task and will be a valuable step in understanding how
common techniques in approximation algorithms can be parallelized
efficiently.  Previous work on facility location commonly relies on
techniques such as linear-program (LP) rounding, local search, primal
dual, and greedy. Unfortunately, LP rounding relies on solving a class
of linear programs not known to be solvable efficiently in
polylogarithmic time.  Neither do known techniques allow for
parallelizing local-search algorithms.  Despite some success in
parallelizing primal-dual and greedy algorithms for set-covering,
vertex-covering, and related problems, these algorithm are obtained
using problem-specific techniques, which are not readily applicable to
other problems.

\smallskip
\subsection{Summary of Results.}
In this paper, we design and analyze several algorithms for (metric)
facility location, $k$-median, $k$-means and $k$-center problems,
focusing on parallelizing a diverse set of techniques in approximation
algorithms.  We study the algorithms on the \textsf{EREW}
\textsf{PRAM} and the Parallel Cache Oblivious
model~\cite{BGV:SPAA10}.  The latter model captures memory locality.
We are primarily concerned with minimizing the work (or cache
complexity) while achieving polylogarithmic depth in these models.  We
are less concerned with polylogarithmic factors in the depth since
such measures are not robust across models.  By work, we mean the
total operation count.  All algorithms we develop are in \NC or \RNC,
so they have polylogarithmic depth.

We first present a parallel \RNC algorithm mimicking the greedy
algorithm of Jain et al.~\cite{JainMMSV:JACM03}. This is the most
challenging algorithm to parallelize because the greedy algorithm is
inherently sequential. We show the algorithm gives a
$(3.722+\varepsilon)$-approximation and does $O(m
\log_{1+\varepsilon}^2 m)$ work, which is within a logarithmic factor
of the serial algorithm. Then, we present a simple $\RNC$ algorithm
using the primal-dual approach of Jain and
Vazirani~\cite{JainV:JACM01} which leads to a
$(3+\varepsilon)$-approximation and for input of size $m$ runs in $O(m
\log_{1+\varepsilon} m)$ work, which is the same as the sequential
work.  The sequential algorithm is a $3$-approximation.  Following
that, we present a local-search algorithm for $k$-median and
$k$-means, with approximation factors of $5 + \vareps$ and $81 +
\vareps$, matching the guarantees of the sequential algorithms. For
constant $k$, the algorithm does $O(n^2\log n)$ work, which is the
same as the sequential counterpart.  Furthermore, we present a
$2$-approximation algorithm for $k$-center with $O((n\log n)^2)$ work,
based on the algorithm of Hochbaum and
Shmoys~\cite{HochbaumS:MathOR85}.  Finally, we show a
$O(m\log^2_{1+\vareps}(m))$-work randomized rounding algorithm, which
yields a $(4+\vareps)$-approximation, given an optimal linear-program
solution as input.  The last two algorithms run in work within a
logarithmic factor of the serial algorithm counterparts.

\subsection{Related Work} 
\label{sec:related-work}

Facility-location problems have had a long history. Because of space
consideration, we mention only some of the results here, focusing on
those concerning metric instances. For the (uncapacitated) metric
facility location, the first constant factor approximation was given
by Shmoys et al.~\cite{ShmoysTA:STOC97}, using an LP-rounding
technique, which has subsequently been improved~\cite{Chudak98,
  GuhaKhuller}. A different approach, based on local-search
techniques, has been used to obtain a $3$-approximation~\cite{KPR98,
  AryaGKMMP:siamjc04,GuptaT:arxiv08}. Combinatorial algorithms based
on primal-dual and greedy approaches with constant approximation
factors are also known~\cite{JainMMSV:JACM03, JainV:JACM01, PT03}.
Other approximation algorithms and hardness results have also been
given by~\cite{Sviridenko02, ChuShm03, Byrka07,CG99, MMSV01, MYZ02-FL,
  KPR98,CG99,GuhaKhuller}. An open problem is to close the gap between
the best known approximation factor of $1.5$~\cite{Byrka07} and the
hardness result of $1.463$~\cite{GuhaKhuller}.

The first constant factor approximation for $k$-median problem was
given by Charikar et al.~\cite{CGTS02}, which was subsequently
improved by~\cite{CG99} and \cite{AryaGKMMP:siamjc04}
to the current best factor of $3 + \varepsilon$. 
For $k$-means, constant-factor
approximations are known for this
problem~\cite{JainV:JACM01,GuptaT:arxiv08}; a special case when the
metric space is the Euclidean space has also been
studied~\cite{KMNPSW03}.  For $k$-center, tight bounds are known:
there is a $2$-approximation algorithm due to
\cite{Gonzalez-kcenter,HochShm86}, and this is tight unless
$\textsf{P} = \textsf{NP}$.

The study of parallel approximation algorithms has been slow since the
early 1990s. There are $\RNC$ and $\NC$ parallel approximation
algorithms for set cover~\cite{BergerRS:focs89,
  RajagopalanV:siamjc98}, vertex
cover~\cite{KhullerVY:jalg94,KoufogiannakisY09:PODC09}, special cases
of linear programs (e.g., positive LPs and cover-packing
LPs)~\cite{LubyN:STOC93,Srinivasan:SODA01,Young01:FOCS01}, and
$k$-center~\cite{WangC:PDP90}. These algorithms are typically based on
parallelizing their sequential counterparts, which usually contain an
inherently sequential component (e.g., a greedy step which requires
picking and processing the minimum-cost element before proceeding to
the next). A common idea in these parallel algorithms is that instead
of picking only the most cost-effective element, they make room for
parallelism by allowing a small slack (e.g., a $(1+\vareps)$ factor)
in what can be selected.  This idea often results in a slightly worse
approximation factor than the sequential version.  For instance, the
parallel set-cover algorithm of Rajagopalan and Vazirani is a
$(2(1+\vareps)\ln n)$-approximation, compared to a $(\ln
n)$-approximation produced by the standard greedy set cover.
Likewise, the parallel vertex-cover algorithm of Khuller et al. is a
$2/(1-\vareps)$-approximation as opposed to the optimal
$2$-approximation given by various known sequential algorithms. Only
recently has the approximation factor for vertex cover been improved
to $2$ in the parallel case~\cite{KoufogiannakisY09:PODC09}.

Several approximation algorithms have been proposed for distributed
computing; see, e.g.~\cite{Elkin:sigact04}, for a survey.  For
facility location, recent research has proposed a number of
algorithms, both for the metric and non-metric
cases~\cite{MoscibrodaW:podc05,GehweilerLS:SPAA06,PanditP:podc09}.
The work of Pandit and Pemmaraju~\cite{PanditP:podc09} is closely
related our primal-dual algorithm; their algorithm is a
$7$-approximation in the {\sf CONGEST} model for distributed
computing.  Both their algorithm and ours have a similar preprocessing
step and rely on the $(1+\vareps)$-slack idea although their algorithm
uses a fixed $\varepsilon = 1$. The model and the efficiency metrics
studied are different, however.



\section{Preliminaries and Notation}
\label{sec:prel-notat}
Let $F$ denote a set of \emph{facilities} and $C$ denote a set of
\emph{clients}. For convenience, let $n_c = |C|$, $n_f = |F|$, and $m
= n_c \times n_f$. Each facility $i \in F$ has a cost of $f_i$, and
each client $j \in C$ incurs a cost (``distance'') $d(j, i)$ to use
the facility $i$. We assume throughout that there is a metric space
$(X, d)$ with $F \cup C \subseteq X$ that underlies our problem
instances. Thus, the distance $d$ is symmetric and satisfies the
triangle inequality.  As a shorthand, denote the cost of the optimal
solution by $\opt$, the facility set of the optimal solution by $F^*$,
and the facility set produced by our algorithm by $F_A$.  Furthermore,
we write $d(u, S)$ to mean the minimum distance from $u$ to a member
of $S$, i.e., $d(u, S) = \min \{ d(u, w) \/:\/ w \in S\}$.

Let $G$ be a graph. We denote by $\deg_G(v)$ the degree of the node
$v$ in $G$ and use $\Gamma_G(v)$ to denote the neighbor set of the
node $v$.  We drop the subscript (i.e., writing $\deg(v)$ and
$\Gamma(v)$) when the context is clear.  Let $V(G)$ and $E(G)$ denote
respectively the set of nodes and the set of edges.

\smallskip
\noindent\textbf{Parallel Models.} 
All the parallel algorithms in this paper can be expressed in terms of
a set of simple operations on vectors and dense matrices, making it
easy to analyze costs on a variety of parallel models.  In particular,
the distances $d(\cdot, \cdot)$ can be represented as a dense $n
\times n$ matrix, where $n = n_c + n_f$, and any data at clients or
facilities can be represented as vectors.  The only operations we need
are parallel loops over the elements of the vector or matrix,
transposing the matrix, sorting the rows of a matrix, and summation,
prefix sums and distribution across the rows or columns of a matrix or
vector.  A prefix sum returns to each element of a sequence the sum of
previous elements.  The summation or prefix sum needs to be applied
using a variety of associative operators, including $\min$, $\max$,
and addition.

We refer to all the operations other than sorting as the \emph{basic
  matrix operation}.  The basic matrix operations on $m$ elements can
all be implemented with $O(m)$ work and $O(\log m)$ time on the
\textsf{EREW} \textsf{PRAM}~\cite{JaJa:book92}, and with $O(m/B)$
cache complexity and $O(\log m)$ depth in the parallel cache oblivious
model.
For the parallel cache oblivious model, we
assume a tall cache $M > B^2$, where $M$ is the size of the cache and
$B$ is the block size.  
Sorting $m$ elements takes $O(m \log m)$ work and $O(\log m)$ time on
an \textsf{EREW} \textsf{PRAM}~\cite{Cole:siamjc88}, and
$O(\frac{m}{B} \log_{M/B} m)$ cache complexity and $O(\log^2 m)$ depth
on the parallel cache oblivious model~\cite{BGV:SPAA10}.  
All algorithms described in this paper are \emph{cache efficient} in
the sense that the cache complexity in the cache oblivious model is
bounded by $O(w/B)$ where $w$ is the work in the EREW model.  All
algorithms use a polylogarithmic number of calls to the basic matrix
operations and sorting and are thus in $\RNC$---do polynomial work
with polylogarithmic depth and possibly use randomization.


Given this set up, the problems considered in this paper can be
defined as follows:
\medskip

\noindent\textbf{(Metric) Facility Location.} The goal of this problem is to find a
set of facilities $F_S \subseteq F$ that minimizes the objective
function
\begin{equation}
  \facloc(F_S) = \sum_{i \in F_S}  f_i + \sum_{j \in C} d(j, F_S)
\end{equation}
Note that we do not need an explicit client-to-facility assignment
because given a set of facilities $F_S$, the cost is minimized by
assigning each client to the closest open facility.

Non-trivial upper- and lower-bounds for the cost of the optimal
solution are useful objects in approximation algorithms. For each
client $j \in C$, let $\gamma_j = \min_{i \in F} (f_i + d(j, i))$ and
$\gamma = \max_{j \in C} \gamma_j$.  The following bounds can be
easily established:
\begin{equation}
  \label{eq:gamma-bounds}
  \gamma \leq \opt \leq \sum_{j \in C} \gamma_j \leq \gamma n_c. 
\end{equation}

Furthermore, metric facility location has a natural integer-program
formulation for which the relaxation yields the pair of primal and
dual programs shown in Figure~\ref{fig:primal-dual-program}.

\begin{figure*}[thb]
\begin{center}
  \small
  \begin{tabular}{l  r}
    \toprule
    \begin{minipage}{0.44\textwidth}
      \begin{tabular}{l l}
          \textbf{Minimize} & 
          $\sum_{i \in F, j \in C} d(j,i)x_{ij} \;\; + \;\; \sum_{i \in F} f_iy_i$ \\
          \mbox{} \\ 
          \textbf{Subj. to:} & \hspace{-5mm}
          $\left \{\begin{array}{l c l l}
            \sum_{i \in F} x_{ij} &  \geq  &  1 &  \text{ for } j \in C\\
             y_i  - x_{ij} &  \geq & 0 & \text{ for } i \in F, j \in C \\
             \multicolumn{3}{l}{x_{ij} \geq 0,\; y_i \geq 0}
          \end{array} \right.$
        \end{tabular}
    \end{minipage}
    &
    \begin{minipage}{0.45\textwidth}
      \begin{tabular}{p{.6in} l}
          \textbf{Maximize} & 
          $\sum_{j \in C} \;\; \alpha_j$\\
          \mbox{} \\
          \textbf{Subj. to:} & \hspace{-5mm}
          $\left \{\begin{array}{l c l l}
            \sum_{j \in C} \beta_{ij} &  \leq  &  f_i &  \text{ for } i \in F\\
            \alpha_j  - \beta_{ij} &  \leq & d(j,i) & \text{ for } i \in F, j \in C \\
            \multicolumn{3}{l}{\beta_{ij} \geq 0,\; \alpha_j \geq 0}
          \end{array} \right.$
        \end{tabular}
    \end{minipage}\\
    \bottomrule
  \end{tabular}
\end{center}
  \caption{The primal (left) and dual (right) programs for metric
    (uncapacitated) facility location.}
\label{fig:primal-dual-program}
\end{figure*}

\smallskip

\noindent\textbf{$k$-Median and $k$-Means.} Unlike facility location, the $k$-median objective does not take into consideration
facility costs, instead limiting the number of opened centers
(facilities) to $k$.  Moreover, in these problems, we typically do not
distinguish between facilities and clients; every node is a client,
and every node can be a facility.  Formally, let $V \subseteq X$ be
the set of nodes, and the goal is to find a set of at most $k$ centers
$F_S \subseteq V$ that minimizes the objective $\kmed(F_S) = \sum_{j
  \in V} d(j, F_S)$. Almost identical to $k$-median is the $k$-means
problem with the objective $\kmeans(F_S) = \sum_{j \in C} d^2(j,
F_S)$.

\medskip

\noindent\textbf{$k$-Center.} Another type of facility-location problem
which has a hard limit on the number of facilities to open is
$k$-center.  The $k$-center problem is to find a set of at most $k$
centers $F_S \subseteq V$ that minimizes the objective $\kcenter(F_S)
= \max_{j \in V} d(j, F_S)$.

In these problems, we will use $n$ to denote the size of $V$.


\section{Dominator Set}
\label{sec:dominator-set}
We introduce and study two variants of the maximal independent set
(MIS) problem, which will prove to be useful in nearly all algorithms
described in this work. The first variant, called the \emph{dominator
  set} problem, concerns finding a maximal set $I \subseteq V$ of
nodes from a simple graph $G = (V,E)$ such that none of these nodes
share a common neighbor (neighboring nodes of $G$ cannot both be
selected).  The second variant, called the \emph{$U$-dominator set}
problem, involves finding a maximal set $I \subseteq U$ of the
$U$-side nodes of a bipartite graph $H=(U,V,E)$ such that none of the
nodes have a common $V$-side neighbor.  We denote by $\maxdom(G)$ and
$\udom(H)$ the solutions to these problems, resp.

\smallskip

Both variants can be equivalently formulated in terms of maximal
independent set. The first variant amounts to finding a maximal
independent set on \[ G^2 = (V, \{uw : uw \in E \text{ or } \exists z
\text{ s.t. } uz, zw \in E\}),\] and the second variant a maximal
independent set on \[H' = (U, \{uw : \exists z \in V \text{ s.t. } uz,
zw \in E\}).\] Because of this relationship, on the surface, it may
seem that one could simply compute $G^2$ or $H'$ and run an existing
MIS algorithm.  Unfortunately, computing graphs such as $G^2$ and $H'$
appears to need $O(n^\omega)$ work, where $\omega$ is the
matrix-multiply constant, whereas the na\"ive greedy-like sequential
algorithms for the same problems run in $O(|E|) = O(n^2)$. This
difference makes it unlikely to obtain work efficient algorithms via
this route.

In this section, we develop near work-efficient algorithms for these
problems, bypassing the construction of the intermediate graphs.  The
key idea is to compute a maximal independent set in-place.  Numerous
parallel algorithms are known for maximal independent set, but the
most relevant to us is an algorithm of Luby~\cite{Luby:siamjc86},
which we now sketch.

The input to the algorithm is a graph $G=(V, E)$.  Luby's algorithm
constructs a maximal independent set $I \subseteq V$ by proceeding in
multiple rounds, with each round performing the following computation:
\begin{algo}[h!]
  \begin{enumerate}
    \denselist

  \item For each $i \in V$, \textbf{in parallel}, $\pi(i) = $ a number
    chosen u.a.r. from $\{1, 2, \dots, 2n^4\}$.

  \item Include a node $i$ in the maximal independent set $I$ if
    $\pi(i) < \min \{\pi(j) : j \in \Gamma(i)\}$, where $\Gamma(i)$ is
    the neighborhood of $i$ in $G$.
  \end{enumerate}
  \caption{The select step of Luby's algorithm for maximal independent set.}
  \label{algo:luby-select-step}
\end{algo}

This process is termed the \emph{select step} in Luby's work.
Following the select step, the newly selected nodes, together with
their neighbors, are removed from the graph before moving on to the
next round.

\medskip

\noindent\textbf{Implementing the select step:} We describe how the select step
can be performed in-place for the first variant; the technique applies
to the other variant. We will be simulating running Luby's algorithm
on $G^2$, without generating $G^2$. Since $G^2$ has the same node set
as $G$, step 1 of Algorithm~\ref{algo:luby-select-step} remains
unchanged.  Thus, the crucial computation for the select step is to
determine efficiently, for each node $i$, whether $\pi(i)$ holds the
smallest number among its neighbors in $G^2$, i.e., computing
efficiently the test in step 2.  To accomplish this, we simply pass
the $\pi(i)$ to their neighbors taking a minimum, and then to the
neighbors again taking a minimum.  These can be implemented with a
constant number of basic matrix operations, in particular distribution
and summation using minimum over the rows and columns of the $|V|^2$
matrix.

\begin{lemma}
  Given a graph $G=(V,E)$, a maximal dominator set $I \subseteq V$ can
  be found in expected $O(\log^2 |V|)$ depth and $O(|V|^2\log |V|)$
  work.  Furthermore, given a bipartite graph $G=(U,V,E)$, a maximal
  $U$-dominator set $I \subseteq U$ can be found in expected $O((\log
  |U|)\cdot \max\{\log |U|, \log |V|\} )$ depth and $O(|V||U|\max\{\log
  |U|, \log |V|\})$ work.
\end{lemma}
For sparse matrices, which we do not use in this paper, this can
easily be improved to $O(|E|\log|V|)$ work.



\section{Facility Location: Greedy}
\label{sec:fac-loc-greedy}

The greedy scheme underlies an exceptionally simple algorithm for
facility location, due to Jain et al.~\cite{JainMMSV:JACM03}.  Despite
the simplicity, the algorithm offers one of the best known
approximation guarantees for the problem.  To describe the algorithm,
we will need some definitions.

\begin{definition}[Star, Price, and Maximal Star]
  A \emph{star} $\fstar = (i, C')$ consists of a facility $i$ and a
  subset $C' \subseteq C$. The \emph{price} of $\fstar$ is
  $\price(\fstar) = (f_i + \sum_{j \in C'} d(j,i))/|C'|$. A star
  $\fstar$ is said to be \emph{maximal} if all \emph{strict} super
  sets of $C'$ have a larger price, i.e., for all $C'' \supsetneq C'$,
  $\price((i, C'')) > \price((i, C'))$.
\end{definition}

The greedy algorithm of Jain et al. proceeds as follows:
\begin{quote}
  Until no client remains, pick the cheapest star $(i, C')$, open the
  facility $i$, set $f_i = 0$, remove all clients in $C'$ from the
  instance, and repeat.
\end{quote}
This algorithm has a sequential running time of $O(m\log m)$ and using
techniques known as factor-revealing LP, Jain et al. show that the
algorithm has an approximation factor of
$1.861$~\cite{JainMMSV:JACM03}.  From a parallelization point of view,
the algorithm is highly sequential---at each step, only the
minimum-cost option is chosen, and every subsequent step depends on
the preceding one.  In this section, we describe how to overcome this
sequential nature and obtain an $\RNC$ algorithm inspired by the
greedy algorithm of Jain et al.  We show that the parallel algorithm
is a $(3.722+\vareps)$-approximation.



The key idea to parallelization is that much faster progress will be
made if we allow a small slack in what can be selected in each round;
however, a subselection step is necessary to ensure that facility and
connection costs are properly accounted for.  



\begin{algo*}[tbh]
\medskip
  In rounds, the algorithm performs the following steps until no
  client remains:
\vspace{-5pt}
\begin{enumerate}
\denselist

\item For each facility $i$, \textbf{in parallel}, compute $\fstar_i =
  (i, C^{(i)})$, the lowest-priced maximal star centered at $i$.

\item Let $\tau = \min_{i \in F} \price(\fstar_i)$, and let $I = \{i
  \in F : \price(\fstar_i) \leq \tau(1+\vareps)\}$.

\item Construct a bipartite graph $H = (I, C', \{ij : d(i,j) \leq
  \tau(1+\vareps) \})$, where $C' = \{j \in C : \exists i \in I \text{
    s.t. } d(i,j) \leq \tau(1+\vareps)\}$.

\item \textbf{Facility Subselection:} while $(I \neq \emptyset)$:
  \vspace{-6pt}
  \begin{enumerate}[(a)]
    \denselist
  \item Let $\Pi: I \to \{1, \dots, |I|\}$ be a random permutation of
    $I$.

  \item For each $j \in C'$, let $\varphi_j = \arg\min_{i \in \Gamma_H(j)}
    \Pi(i)$.

  \item For each $i \in I$, if $|\{j: \varphi_j = i\}| \geq
    \frac1{2(1+\vareps)}\deg(i)$, add $i$ to $F_A$ (open $i$), 
    set $f_i = 0$,
    remove $i$ from $I$ , and remove $\Gamma_H(i)$ from both $C$ and
    $C'$.  

    \textit{Note:} \textit{ In the analysis, the clients removed in
      this step have $\pi_j$ set as follows. If the facility
      $\varphi_j$ is opened, let $\pi_j = \varphi_j$; otherwise,
      $\pi_j$ is set to any facility $i$ we open in this step such
      that $ij \in E(H)$.  Note that any facility that is opened is
      at least $1/(2(1+\vareps))$ paid for by the clients that select it, 
      and that since every client
      is assigned to at most one facility, they only pay for one edge.}
  \item Remove $i \in I$ (and the incident edges) from the graph $H$
    if on the remaining graph, $\frac{f_i + \sum_{j \in \Gamma_H(i)}
      d(j,i)}{\deg(i)} > \tau(1+\vareps)$. These facilities will show
    up in the next round (outer-loop).

  \end{enumerate}
  \textit{Note: After $f_i$ is set to $0$, facility $i$ will still
    show up in the next round.}
\end{enumerate}
\caption{Parallel greedy algorithm for metric facility location.}
\label{algo:greedy-main}
\vspace{-2mm}
\end{algo*}

We present the parallel algorithm in Algorithm~\ref{algo:greedy-main}
and now describe step 1 in greater detail; steps $2$ -- $3$ can be
implemented using standard
techniques~\cite{JaJa:book92,Leighton:book92}. As observed in Jain et
al.~\cite{JainMMSV:JACM03} (see also Fact~\ref{fact:star-maximality}),
for each facility $i$, the lowest-priced star centered at $i$ consists
of the $\kappa_i$ closest clients to $i$, for some
$\kappa_i$. Following this observation, we can presort the distance
between facilities and clients for each facility.  Let $i$ be a
facility and assume without loss of generality that $d(i,1) \leq
d(i,2) \leq \dots \leq d(i, n_c)$. Then, the cheapest maximal star for
this facility can be found as follows. Using prefix sum, compute the
sequence $p^{(i)} = \{(f_i + \sum_{j\leq k}
d(i,k))/k\}_{k=1}^{n_c}$. Then, find the smallest index $k$ such that
$p^{(i)}_k < p^{(i)}_{k+1}$ or use $k = n_c$ if no such index
exists. It is easy to see that the maximal lowest-priced star centered
at $i$ is the facility $i$ together with the client set $\{1, \dots,
k\}$.

Crucial to this algorithm is a subselection step, which ensures that
every facility and the clients that connect to it are adequately
accounted for in the dual-fitting analysis.  This subselection process
can be seen as scaling back on the aggressiveness of opening up the
facilities, mimicking the greedy algorithm's behavior more closely.


\subsection{Analysis}

We present a dual-fitting analysis of the above algorithm. The
analysis relies on the client-to-facility assignment $\pi$, defined in
the description of the algorithm.  The following easy-to-check facts
will be useful in the analysis.
\begin{fact}
  \label{fact:star-maximality}
  For each iteration of the execution, the following holds: (1) If
  $\fstar_i$ is the cheapest maximal star centered at $i$, then $j$
  appears in $\fstar_i$ if and only if $d(j, i) \leq
  \price(\fstar_i)$. (2) If $t = \price(\fstar_i)$, then $\sum_{j \in
    C} \max(0, t - d(j,i)) = f_i$.
\end{fact}

Now consider the dual program in Figure~\ref{fig:primal-dual-program}.
For each client $j$, set $\alpha_j$ to be the $\tau$ setting in the
iteration that the client was removed.  
We begin the analysis by relating the cost of the solution that the
algorithm outputs to the cost of the dual program.
\begin{lemma}
  The cost of the algorithm's solution $\sum_{i \in F_A} f_i + \sum_{j
    \in C} d(j,F_A)$ is upper-bounded by $2(1+\vareps)^2\sum_{j \in C}
  \alpha_j$.
\end{lemma}

\begin{proof}
  Consider that in step 4(c), a facility $i$ is opened if at least a
  $\frac1{2(1+\vareps)}$ fraction of the neighbors ``chose'' $i$.
  Furthermore, we know from the definition of $H$ that, in that round,
  $f_i +\sum_{j \in \Gamma_H(i)} d(j,i) \leq \tau(1+\vareps)\deg(i)$.
  By noting that we can partition $C$ by which facility the client is
  assigned to in the assignment $\pi$, we establish
  \begin{align*}
    \sum_{j \in C} \alpha_j\cdot 2(1+\vareps)^2
    &\geq \sum_{i \in F_A}\Big( f_i + \sum_{j: \pi_j =i} d(j, i)\Big)\\
    &\geq  \sum_{i \in F_A} f_i + \sum_{j \in C} d(j,F_A), 
  \end{align*}
  as desired.
\end{proof}

In the series of claims that follows, we show that when scaled down by
a factor of $\gamma = 1.861$, the $\alpha$ setting determined above is
a dual feasible solution. We will assume without loss of generality
that $\alpha_1 \leq \alpha_2 \leq \dots \leq \alpha_{n_c}$. Let $W_i =
\{ j \in C \/:\/ \alpha_j \geq \gamma \cdot d(j,i) \}$ for all $i \in
F$ and $W = \cup_i W_i$.

\begin{claim}
  \label{claim:greedy-fac-cost-bound}
  For any facility $i \in F$ and client $j_0 \in C$,
  \begin{equation*}
    \sum_{j \in W\!: j \geq j_0} \max(0, \alpha_{j_0} - d(j, i)) \;\; \leq \;\; f_i.
  \end{equation*}
\end{claim}
\begin{proof} 
  Suppose for a contradiction that there exist client $j$ and facility
  $i$ such that the inequality in the claim does not hold. That is,
  \begin{equation}
    \sum_{j \in W\!: j \geq j_0} \max(0, \alpha_{j_0} - d(j, i)) \;\;
    > \;\; f_i \label{eq:fac-open-time}.
  \end{equation}
  Consider the iteration in which $\tau$ is $\alpha_{j_0}$; call this
  iteration $\ell$.  By Equation~\eqref{eq:fac-open-time}, there
  exists a client $j \in W \cap \{j \in \Z_+ : j \geq j_0\}$ such that
  $\alpha_{j_0} - d(j,i) > 0$; thus, this client participated in a
  star in an iteration prior to $\ell$ and was connected
  up. Therefore, it must be the case that $\alpha_j < \alpha_{j_0}$,
  which is a contradiction to our assumption that $j_0 \leq j$ and
  $\alpha_1 \leq \alpha_2 \leq .. \dots \alpha_{n_c}$.
\end{proof}

\begin{claim}
  \label{claim:greedy-rel-cost}
  Let $i \in F$, and $j, j' \in W$ be clients. Then, $\alpha_j \leq
  \alpha_{j'} + d(i,j') + d(i,j)$.
\end{claim}

The proof of this claim closely parallels that of Jain et
al.~\cite{JainMMSV:JACM03} and is omitted.  These two claims form the
basis for the set up of Jain et al.'s factor-revealing LP.  Hence,
combining them with Lemmas 3.4 and 3.6 of Jain et
al.~\cite{JainMMSV:JACM03}, we have the following lemma:

\begin{lemma}
  The setting $\alpha'_j = \frac{\alpha_j}{\gamma}$ and $\beta'_{ij} =
  \max(0,\alpha'_j - d(j,i))$ is a dual feasible solution, where
  $\gamma = 1.861$.
\end{lemma}

\noindent\textbf{An Alternative Proof Without Factor-Revealing LP.}
We note that a slightly weaker result can be derived without the use
of factor-revealing LP.  Claims~\ref{claim:greedy-rel-cost}
and~\ref{claim:greedy-fac-cost-bound} can be combined to prove the
following lemma:

\begin{lemma}
  The setting $\alpha'_j = {\alpha_j}/{3}$ and $\beta'_{ij} =
  \max(0,\alpha'_j - d(j,i))$ is a dual feasible solution.
\end{lemma}
\begin{proof}
  We will show that for each facility $i \in F$,
  \begin{equation}
    \sum_{j \in W_i} \big(\alpha_j - 3\cdot d(j,i)\big) \;\;\leq\;\; 3\cdot f_i.
    \label{eq:greedy-contrib}
  \end{equation}
  Note that if $W_i$ is empty, the lemma is trivially true. Thus, we
  assume $W_i$ is non-empty and define $j_0$ to be $\min W_i$. Since
  $j_0 \in W_i$, $d(j_0,i) \leq \alpha_{j_0}$ by the definition of
  $W_i$. Now let $T = \{j \in W_i : \alpha_{j_0} \geq d(j,i)
  \}$. Applying Claims~\ref{claim:greedy-rel-cost}
  and~\ref{claim:greedy-fac-cost-bound}, we have
  \begin{align*}
    \sum_{j \in W_i} (\alpha_j - d(j,i)) 
    &\leq \sum_{j \in W_i} (\alpha_{j_0} + d(j_0,i))
    \leq \sum_{j\in W_i} 2\cdot \alpha_{j_0} \\
    &\leq 2f_i + \sum_{j \in T} 2\cdot d(j,i)
     + \sum_{j \in W_i\setminus T} 2\cdot d(j,i)  
     \leq 2f_i + \sum_{j \in W_i} 2\cdot d(j,i),
  \end{align*}
  which proves inequality~\eqref{eq:greedy-contrib}.  With this, it is
  easy to see that our choice of $\beta'_{ij}$'s ensures that all
  constraints of the form $\alpha_j - \beta_{ij} \leq d(j,i)$ are
  satisfied.  Then, by inequality~\eqref{eq:greedy-contrib}, we have
  $\sum_{j \in C} \max(0, \alpha_j - 3\cdot d(j,i)) = \sum_{j \in W_i}
  [\alpha_j - 3\cdot d(j,i)] \leq 3 \cdot f_i$, which implies that
  $\sum_{j \in C} \max(0, \alpha_j - 3\cdot d(j,i)) \leq 3 \cdot f_i$.
  Hence, we conclude that for all facility $i \in F$, $\sum_{j \in C}
  \beta'_{ij} \leq f_i$, proving the lemma.
\end{proof}

\paragraph{Running time analysis} 
\smallskip
Consider the algorithm's description in
Algorithm~\ref{algo:greedy-main}.  The rows can be presorted to give
each client its distances from facilities in order.  In the original
order, each element can be marked with its rank.  Step 1 then involves
a prefix sum on the sorted order to determine how far down the order
to go and then selection of all facilities at or below that rank.
Steps 2--3 require reductions and distributions across the rows or
columns of the matrix.  The subset $I \subset F$ can be represented as
a bit mask over $F$.  Step 4 is more interesting to analyze; the
following lemma bounds the number of rounds facility subselection is
executed, the proof of which is analogous to Lemma~4.1.2 of
Rajagopalan and Vazirani~\cite{RajagopalanV:siamjc98}; we present here
for completeness a simplified version of their proof, which suffices
for our lemma.

\begin{lemma}
  \label{lem:greedy-subselection-rounds}
  With probability $1 - o(1)$, the subselection step terminates within
  $O(\log_{1+\vareps} m)$ rounds.
\end{lemma}

\begin{proof}
  Let $\Phi = |E|$. We will show that if $\Phi'$ is the potential
  value after an iteration of the subselection step, then $\expct{\Phi
    - \Phi'} \geq c\Phi$, for some constant $c > 0$. The lemma then
  follows from standard results in probability theory.  To proceed,
  define ${\sf chosen}_i = |\{ j \in C' : \varphi_j = i\}|$.
  Furthermore, we say that an edge $ij$ is \emph{good} if at most
  $\theta = \frac12(1 - \frac1{1+\vareps})$ fraction of neighbors of
  $i$ have degree higher than $j$.

  Consider a good edge $ij$. We will estimate $\expct{{\sf chosen}_i |
    \varphi_j = i}$. Since $ij$ is good, we know that \[ \sum_{j' \in
    \Gamma_H(i)} \bm{1}_{\{\deg(j') \leq \deg(j)\}} \geq (1 -
  \theta)\deg(i).\] Therefore, $\expct{{\sf chosen}_i | \varphi_j =i }
  \geq \frac12(1 - \theta)\deg(i)$, as it can be shown that
  $\prob{\varphi_{j'} = i | \varphi_{j} = i} \geq \frac12$ for all $j'
  \in \Gamma_H(i)$ and $\deg(j') \leq \deg(j)$.  By Markov's
  inequality and realizing that ${\sf chosen}_i \leq \deg(i)$, we have
  \[\prob{ {\sf chosen}_i \geq \frac{1}{2(1+\vareps)}\deg(i) \;\;\Big |\;\;
    \varphi_j = i} = p_0 > 0.\]
  Finally, we note that $\expct{\Phi - \Phi'}$ is at least 
  \begin{align*}
    &\sum_{ij \in E} \prob{\varphi_j = i \text{ and } {\sf chosen}_i
      \geq \frac{1}{2(1 + \vareps)}\deg(i) }\cdot \deg(j) \\
    &\;\;\;\geq
    \sum_{\textrm{good } ij \in E} \frac{1}{\deg(j)} p_0 \deg(j) \\
    &\;\;\;\geq
    p_0\sum_{ij \in E} \bm{1}_{\{ij \text{ is good}\}}.
  \end{align*}
  Since at least $\theta$ fraction of the edges are good, $\expct{\Phi
    - \Phi'} \geq p_0\theta \Phi$. Since $\ln (1/(1 - p_0\theta)) =
  \Omega(\log (1 + \vareps))$, the lemma follows from standard results
  in probability~\cite{MotwaniR:book95}.
\end{proof}

It is easy to see that each subselection step can be performed with a
constant number of basic matrix operations over the $D$ matrix.
Therefore, if the number of rounds the main body is executed is $r$,
the algorithm makes $O(r\log_{1+\vareps} m)$ calls to the basic matrix
operations described in Section~\ref{sec:prel-notat} with probability
exceeding $1 - o(1)$.  It also requires a single sort in the
preprocessing.  This means $O(r\log_{1+\vareps} m \log m)$ time
implies a total of $O(rm\log_{1+\vareps} m)$ work (with probability
exceeding $1 - o(1)$) on the \textsf{EREW} \textsf{PRAM}.
Furthermore, it is cache efficient (cache complexity is $O(w/B)$)
since the sort is only applied once and does not dominate the cache
bounds.

\paragraph{Bounding the number of rounds}
\smallskip

Before describing a less restrictive alternative, we point out that
the simplest way to bound the number of rounds by a polylogarithm
factor is to rely on the common assumption that the facility cost, as
well as the ratio between the minimum (non-zero) and the maximum
client-facility distance, is polynomially bounded in the input
size. As a result of this assumption, the number of rounds is
upper-bounded by $\log_{1+\vareps} (m^c) = O(\log_{1+\vareps} m)$, for
some $c \geq 1$.

Alternatively, we can apply a preprocessing step to ensure that the
number of rounds is polylogarithm in $m$.  The basic idea of the
preprocessing step is that if a star is ``relatively cheap,''
opening it right away will harm the approximation factor only
slightly.
%
Using the bounds in Equation~(\ref{eq:gamma-bounds}), if
$\fstar_i$ is the lowest-priced maximal star centered at $i$, we know
we can afford to open $i$ and discard all clients attached to it if
$\price(\fstar_i) \leq \frac{\gamma}{m^2}$.  Therefore, the
preprocessing step involves: (1) computing $\fstar_i$, the
lowest-priced maximal star centered at $i$, for all $i \in F$, (2)
opening all $i$ such that $\price(\fstar_i) \leq \frac{\gamma}{m^2}$,
(3) setting $f_i$ of these facilities to $0$ and removing all clients
attached to these facilities.

Computing $\gamma$ takes $O(\log n_c + \log n_f)$ depth and $O(m)$
work. The rest of the preprocessing step is at most as costly as a
step in the main body.  Thus, the whole preprocessing step can be
accomplished in $O(\log m)$ depth and $O(m)$ work.  With this
preprocessing step, three things are clear: First, $\tau$ in the first
iteration of the main algorithm will be at least $\frac{\gamma}{m^2}$,
because cheaper stars have already been processed in preprocessing.
Second, the cost of our final solution is increased by at most $n_c
\times \frac{\gamma}{m^2} \leq \frac{\gamma}{m} \leq \opt/m$, because
the facilities and clients handled in preprocessing can be accounted
for by the cost of their corresponding stars---specifically, there can
be most $n_c$ stars handled in preprocessing, each of which has price
$\leq \gamma/m^2$; and the price for a star includes both the facility
cost and the connection cost of the relevant clients and
facilities. Finally, in the final iteration, $\tau \leq n_c\gamma$.
As a direct consequence of these observations, the number of rounds is
upper-bounded by $\log_{1+\vareps} (\frac{n_c\gamma}{\gamma/m^2}) \leq
\log_{1+\vareps} (m^3) = O(\log_{1+\vareps} m)$, culminating in the
following theorem:

\begin{theorem}
  Let $0 < \vareps \leq 1$ be fixed. For sufficiently large input,
  there is a greedy-style $\RNC$ $O(m\log^2_{1+\vareps} (m))$-work algorithm that
  yields a factor-$(6+\vareps)$ approximation for the metric
  facility-location problem.
\end{theorem}



\section{Facility Location: Primal-Dual}
\label{sec:fac-loc-primal-dual}

The primal-dual scheme is a versatile paradigm for combinatorial
algorithms design. In the context of facility location, this scheme
underlies the Lagrangian-multiplier preserving\footnote{This means
  $\alpha\sum_{i \in F_A} f_i + \sum_{j \in C} d(j, F_A) \leq \alpha
  \cdot \opt$, where $\alpha$ is the approximation ratio.} (LMP)
$3$-approximation algorithm of Jain and Vazirani, enabling them to use
the algorithm as a subroutine in their $6$-approximation algorithm for
$k$-median~\cite{JainV:JACM01}.

The algorithm of Jain and Vazirani consists of two phases, a
primal-dual phase and a postprocessing phase. To summarize this
algorithm, consider the primal and dual programs in
Figure~\ref{fig:primal-dual-program}. In the primal-dual phase,
starting with all dual variables set to $0$, we raise the dual
variables $\alpha_j$'s uniformly until a constraint of the form
$\alpha_j - \beta_{ij} \leq d(j, i)$ becomes tight, at which point
$\beta_{ij}$ will also be raised, again, uniformly to prevent these
constraints from becoming overtight.  When a constraint $\sum_{j}
\beta_{ij} \leq f_i$ is tight, facility $i$ is tentatively opened and
clients with $\alpha_j \geq d(j,i)$ are ``frozen,'' i.e., we stop
raising their $\alpha_j$ values from this point on.  The first phase
ends when all clients are frozen.  In the postprocessing phase, we
compute and output a maximal independent set on a graph $G$ of
tentatively open facilities; in this graph, there is an edge between a
pair of facilities $i$ and $i'$ if there is a client $j$ such that
$\alpha_j > d(j,i)$ and $\alpha_j > d(j,i')$.  Thus, the maximal
independent set ensures proper accounting of the facility cost (i.e.,
each client ``contributes'' to at most one open facility, and every
open facility has enough contribution).  Informally, we say that a
client $j$ ``pays'' for or ``contributes'' to a facility $i$ if
$\beta_{ij} = \alpha_j - d(j,i) > 0$.
\smallskip

\noindent\emph{Remarks.} We note that in the parallel setting, the
description of the postprocessing step above does not directly lead to
an efficient algorithm, because constructing $G$ in polylogarithmic
depth seems to need $O(m n_f)$ work, which is much more than one needs
sequentially.
\smallskip

In this section, we show how to obtain a work-efficient $\RNC$ $(3 +
\vareps)$-approximation algorithm for facility location, based on the
primal-dual algorithm of Jain and Vazirani. 
%
%
Critical to bounding the number of iterations in the main algorithm by
$O(\log m)$ is a preprocessing step, which is similar to that used by
Pandit and Pemmaraju in their distributed
algorithm~\cite{PanditP:podc09}.

%
\smallskip
\noindent\textbf{Preprocessing:} Assuming $\gamma$ as defined in
Equation~(\ref{eq:gamma-bounds}), we will open every facility $i$ that
satisfies \[ \sum_{j \in C} \max\left(0, \frac{\gamma}{m^2} - d(j,
  i)\right) \;\; \geq \;\; f_i. \] Furthermore, for all clients $j$
such that there exists an opened $i$ and $d(j, i) \leq \gamma/m^2$, we
declare them connected and set $\alpha_j = 0$.  The facilities opened
in this step will be called \emph{free} facilities and denoted by the
set $F_0$.

\smallskip
\noindent\textbf{Main Algorithm:} The main body of the algorithm is described
in Algorithm~\ref{algo:primal-dual}. The algorithm outputs a bipartite
graph $H = (F_T, C, E)$, constructed as the algorithm executes.  Here
$F_T$ is the set of facilities declared open during the iterations of
the main algorithm and $E$ is given by $E = \{ij : i \in F, j \in C,
\text{ and } (1+\vareps)\alpha_j > d(j, i)\}$.

\begin{algo*}[tbh]
  \medskip For iteration $\ell=0, 1, \dots$, the algorithm performs
  the following steps until all facilities are opened or all clients
  are frozen, whichever happens first.  
  \vspace{-2mm}
  \begin{enumerate}
    \denselist
  \item For each unfrozen client $j$, \textbf{in parallel}, set
    $\alpha_j$ to $\frac{\gamma}{m^2}(1+\vareps)^{\ell}$.
    
  \item For each unopened facility $i$, \textbf{in parallel}, declare it
    open if
    \begin{equation*}
      \sum_{j \in C} \max(0, (1 + \vareps)\alpha_j - d(j,i)) \geq f_i.
    \end{equation*}
    
  \item For each unfrozen client $j$, \textbf{in parallel}, freeze this
    client if there exists an opened facility $i$ such that
    $(1+\vareps)\alpha_j \geq d(j, i)$.
  \item Update the graph $H$ by adding edges between pairs of nodes
    $ij$ such that $(1+\vareps)\alpha_j > d(j,i)$.
  \end{enumerate}
  After the last iteration, if all facilities are opened but some
  clients are \emph{not} yet frozen, we determine in parallel the
  $\alpha_j$ settings of these clients that will make them reach an
  open facility (i.e., $\alpha_j = \min_i d(j, i)$). Finally, update
  the graph $H$ as necessary.
  \caption{Parallel primal-dual algorithm for metric facility location}
  \label{algo:primal-dual}
\end{algo*}

\smallskip

\noindent
\textbf{Post-processing.} As a post-processing step, we compute $I =
\udom(H)$. Thus, the set of facilities $I \subseteq F_T$ has the
property that each client contributes to the cost of at most one
facility in $I$. Finally, we report $F_A = I \cup F_0$ as the set of
facilities in the final solution.

\subsection{Analysis}

To analyze approximation guarantee of this algorithm, we start by
establishing that the $\alpha_j$ setting produced by the algorithm
leads to a dual feasible solution.

\begin{claim}
  \label{claim:primal-dual-incr}
  For any facility $i$, \[\sum_{j \in \Gamma_H(i)} \max(0, \alpha_j -
  d(j,i)) \leq f_i.\]
\end{claim}
\begin{proof}
  Let $\alpha_j^{(\ell)}$ denote the $\alpha_j$ value at the end of
  iteration $\ell$. Suppose for a contradiction that there is a
  facility $i$ which is overtight. More formally, there exists $i \in
  F$ and the smallest $\ell$ such that $\sum_{j \in \Gamma_F(i)}
  \max(0, \alpha_j^{(\ell)} - d(j,i)) > f_i$.  Let $J$ be the set of
  unfrozen neighboring clients of $i$ in iteration $\ell - 1$. The
  reason facility $i$ was not opened in iteration $\ell - 1$ and the
  surrounding clients were not frozen is
   \begin{equation*}
     \textit{raised}_i \stackrel{{\sf def}}{=} 
     \sum_{j \in \Gamma_F(i)\setminus J} \max(0,
     (1+\vareps)\alpha_j^{(\ell-1)} - d(j,i)) + \sum_{j \in J} \max(0,
     (1+\vareps)t_{\ell-1} - d(j,i)) < f_i.
   \end{equation*}
   However, we know that $t_{\ell} = (1+\vareps)t_{\ell-1}$, and for
   each frozen neighboring client $j$ (i.e., $j \in \Gamma_F(i)
   \setminus J$), $\alpha_j^{(\ell)} = \alpha_j^{(\ell-1)}$, so
   \begin{equation*}
     \textit{raised}_i \geq \sum_{j \in \Gamma(i)\setminus J} \max(0,
     \alpha_j^{(\ell)} - d(j,i)) + \sum_{j \in J} \max(0,
     t_\ell - d(j,i)) = \sum_{j \in \Gamma(i)} \max(0, \alpha_j^{(\ell)} - d(j,i)),
   \end{equation*}
   which is a contradiction.
\end{proof}

It follows from this claim that setting $\beta_{ij} = \max(0, \alpha_j
- d(j,i))$ provides a dual feasible solution. Next we relate the cost
of our solution to the cost of the dual solution. To ease the
following analyses, we use a client-to-facility assignment $\pi\!: C
\to F$, defined as follows: For all $j \in C$, let $\varphi(j) = \{i :
(1+\vareps)\alpha_j \geq d(j, i)\}$.  Now for each client $j$,
\textbf{(1)} if there exists $i \in F_0$ such that $d(j, i) \leq
\gamma/m^2$, set $\pi_j$ to \emph{any} such $i$; \textbf{(2)} if there
exists $i \in I$ such that $ij$ is an edge in $H$, then $\pi_j = i$
($i$ is unique because of properties of $I$) ; \textbf{(3)} if there
exists $i \in I$ such that $i \in \varphi(j)$, then $\pi_j = i$;
\textbf{(4)} otherwise, pick $i' \in \varphi(j)$ and set $\pi_j$ to $i
\in I$ which is a neighbor of a neighbor of $i'$.

Clients of the first case, denoted by $C_0$, are called \emph{freely
  connected}; clients of the cases (2) and (3), denoted by $C_1$, are
called \emph{directly connected}. Otherwise, a client is
\emph{indirectly connected}.


The following lemmas bound the facility costs and the connection costs
of indirectly connected clients.

\begin{lemma}
\label{lem:primal-dual-fac-cost}
\begin{equation*}
  \sum_{i \in F_A} f_i \;\; \leq \;\;  \frac{\gamma}m  \; +  \;
 \sum_{j \in C_1 } (1+\vareps)\alpha_j - \sum_{j \in C_0 \cup C_1} d(j,\pi_j)
\end{equation*}
\end{lemma}

\begin{proof}
  When facility $i \in F_T$ was opened, it must satisfy $f_i \leq
  \sum_{j: ij \in E(G)} (1+\vareps)\alpha_j - d(j,i)$.  If client $j$
  has contributed to $i$ (i.e., $(1+\vareps)\alpha_j - d(j,i) > 0$)
  and $i \in I$, then $j$ is directly connected to it. Furthermore,
  for each client $j$, there is at most one facility in $I$ that it
  contributes to (because $I =\udom(H)$). Therefore, $\sum_{i \in I}
  f_i \leq \sum_{j \in C_1 } (1+\vareps)\alpha_j -
  d(j,\pi_j)$. Furthermore, for each ``free'' facility, we know that
  $f_i \leq \sum_{j \in C} \max(0, \gamma^2/m^2 - d(j, i))$, so by our
  choice of $\pi$, $ f_i \leq \frac{\gamma}{m^2}\times n_c - \sum_{j
    \in C_0: \pi_j = i} d(j, i)$.  Thus, $\sum_{i \in F_0} f_i \leq
  \gamma/m - \sum_{j \in C_0} d(j,i)$.  Combining these results and
  observing that $F_A$ is the disjoint union of $I$ and $F_0$, we have
  the lemma.
\end{proof}

\begin{lemma}
\label{lem:primal-dual-indir-conn-cost}
  For each indirectly connected client $j$ (i.e., $j \not\in C_0 \cup
  C_1$), we have $d(j, \pi_j) \leq 3(1 + \vareps)\alpha_j$.
\end{lemma}

\begin{proof}
  Because $j \not\in C_0 \cup C_1$ and $I = \udom(H)$, there must
  exist a facility $i' \in \varphi(j)$ and a client $j'$ such that
  $j'$ contributed to both $i$ and $i'$, and $(1+\vareps)\alpha_j \geq
  d(j,i')$. We claim that both $d(j',i')$ and $d(j',i)$ are
  upper-bounded by $(1+\vareps)\alpha_j$. To see this, we note that
  because $j'$ contributed to both $i$ and $i'$, $d(j',i') \leq
  (1+\vareps)\alpha_{j'}$ and $d(j',i) \leq (1+\vareps)\alpha_{j'}$.
  Let $\ell$ be the iteration in which $j$ was declared frozen, so
  $\alpha_j = t_\ell$. Since $i' \in \varphi(j)$, $i'$ must be
  declared open in iteration $\leq \ell$.  Furthermore, because
  $(1+\vareps)\alpha_{j'} > d(j',i')$, $\alpha_{j'}$ must be frozen in
  or prior to iteration $\ell$.  Consequently, we have $\alpha_{j'}
  \leq t_\ell = \alpha_j$.  Combining these facts and applying the
  triangle inequality, we get $d(j,i) \leq d(j,i') + d(i',j') +
  d(j',i) \leq (1+\vareps)\alpha_j + 2(1+\vareps)\alpha_{j'} \leq
  3(1+\vareps)\alpha_j$.
\end{proof}

By
Lemmas~\ref{lem:primal-dual-fac-cost}~and~\ref{lem:primal-dual-indir-conn-cost},
we establish
\begin{equation}
  3\sum_{i \in F_A} f_i + \sum_{j \in C} d(j,\pi_j) \;\;\leq\;\;  
  \frac{3\gamma}{m} + 3(1+\vareps)\sum_{j \in C} \alpha_j.
  \label{eq:totalcost-dual}
\end{equation}

Now since $\{\alpha_j,\beta_{ij}\}$ is dual feasible, its value can
be at most that of the primal optimal solution; that is, $\sum_j
\alpha_j \leq \opt$.  Therefore, combining with
Equation~\eqref{eq:totalcost-dual}, we know that the cost of the
solution returned by parallel primal-dual algorithm in this section is
at most $ 3\sum_{i \in F_A} f_i + \sum_{j \in C} d(j,C) \leq
(3+\vareps')\opt$ for some $\vareps' > 0$ when the problem instance is
large enough.


\paragraph{Running Time Analysis}
\smallskip
We analyze the running of the algorithm presented, starting with the
main body of the algorithm.  Since $\sum_j \alpha_j \leq \opt$ and
$\opt \leq n_c \gamma$, no $\alpha_j$ can be bigger than $n_c\gamma \leq
m\gamma$. Hence, the main algorithm must terminate before $\ell >
3\log_{1+\vareps} m$, which upper-bounds the number of iterations to
$O(\log_{1+\vareps} m)$.  In each iteration, steps 1, 3, and 4 perform
trivial work. Step 2 can be broken down into (1) computing the $\max$
for all $i \in F, j \in C$,, and (2) computing the sum for each $i
\in F$.  These can all be implemented with the basic matrix operations,
giving a totalof $O(\log_{1+\vareps} m)$ of basic matrix operations
over a matrix of size $m$.

The preprocessing step, again, involves some reductions over the rows
and columns of the matrix.  This includes the calculations of
$\gamma_j$'s and the composite $\gamma$.  The post-processing step
relies on computing the $U$-dominating set, as described in
Section~\ref{sec:prel-notat} which runs in $O(\log m)$ matrix
operations.

The whole algorithm therefore runs in $O(\log_{1+\vareps} m)$ basic matrix
operations and is hence work efficient compared to the $O(m\log m)$
sequential algorithm of Jain and Vazirani.  Putting these altogether,
we have the following theorem:

\begin{theorem}
  Let $\vareps > 0$ be fixed.  For sufficiently large $m$, there is a
  primal-dual $\RNC$ $O(m\log_{1+\vareps} m)$-work algorithm that
  yields a factor-$(3+\vareps)$ approximation for the metric
  facility-location problem.
\end{theorem}


\section{Other Results}
\label{sec:other-results}
In this section, we consider other applications of dominator set in
facility-location problems.

\subsection{\texorpdfstring{$k$-Center}{k-Center}}
\label{sec:k-center}

Hochbaum and Shmoys~\cite{HochbaumS:MathOR85} show a simple factor-$2$
approximation for $k$-center. The algorithm performs a binary search
on the range of distances. We show how to combine the dominator-set
algorithm from Section~\ref{sec:dominator-set} with standard
techniques to parallelize the algorithm of Hochbaum and Shmoys,
resulting in an $\RNC$ algorithm with the same approximation
guarantee.  Consider the set of distances $\mathcal{D} = \{d(i, j) : i
\in C \text{ and } j \in V\}$ and order them so that $d_1 < d_2 <
\dots < d_{p}$ and $\{d_1,\dots, d_{p}\} = \mathcal{D}$, where $p =
|\mathcal{D}|$. The sequence $\{d_i\}_{i=1}^p$ can be computed in
$O(\log |V|)$ depth and $O(|V|^2 \log |V|)$ work.  Let $H_\alpha$ be a
graph defined as follows: the nodes of $H_\alpha$ is the set of nodes
$V$, but there is an edge connecting $i$ and $j$ if and only if $d(i,
j) \leq \alpha$.

The main idea of the algorithm is simple: find the smallest index $t
\in \{1, 2, \dots, p\}$ such that $\maxdom(H_{d_t}) \leq k$. Hochbaum
and Shmoys observe that the value $t$ can be found using binary search
in $O(\log p) = O(\log |V|)$ probes. We parallelize the probe step,
consisting of constructing $H_{d_{t'}}$ for a given $t' \in \{1,
\dots, p\}$ and checking whether $|\maxdom(H_{d_{t'}})|$ is bigger
than $k$.  Constructing $H_{d_{t'}}$ takes $O(1)$ depth and $O(|V|^2)$
work, and using the maximal-dominator-set algorithm from
Section~\ref{sec:dominator-set}, the test can be performed in expected
$O(\log^2 |V|)$ depth and expected $O(|V|^2\log|V|)$ work.  The
approximation factor is identical to the original algorithm, hence
proving the following theorem:

\begin{theorem}
  There is an $\RNC$ $2$-approximation algorithm with
  $O((|V|\log|V|)^2)$~work for $k$-center.
\end{theorem}

\subsection{Facility Location: LP Rounding}
\label{sec:fac-loc-lp-rounding}

LP rounding was among the very first techniques that yield non-trivial
approximation guarantees for metric facility location. The first
constant-approximation algorithm was given by Shmoys et
al.~\cite{ShmoysTA:STOC97}.  Although we do not know how to solve the
linear program for facility location in polylogarithmic depth, we
demonstrate another application of the dominator-set algorithm and the
slack idea by parallelizing the randomized-rounding step of Shmoys et
al. The algorithm yields a $(4+\vareps)$-approximation, and the
randomized rounding is an \RNC algorithm.

The randomized rounding algorithm of Shmoys et al.~consists of two
phases: a filtering phase and a rounding phase. In the following, we
show how to parallelize these phases and prove that the parallel
version has a similar guarantee. Our presentation differs slightly
from the original work but works in the same spirit.

\medskip
\noindent\textbf{Filtering:} The filtering phase is naturally parallelizable.
Fix $\alpha$ to be a value between $0$ and $1$. Given an optimal
primal solution $(x, y)$, the goal of this step is to produce a new
solution $(x' ,y')$ with properties as detailed in
Lemma~\ref{lemma:lp-rounding-filtered-cost}. Let $\delta_j = \sum_{i
  \in F} d(i,j)\cdot x_{ij}$, $B_j = \{ i \in F : d(i, j) \leq
(1+\alpha)\delta_j\}$, and ${\sf mass}(B_j)= \sum_{i \in B_j} x_{ij}$.
We compute $x'_{ij}$ and $y'_i$ as follows: (1) let $x'_{ij} =
{x_{ij}}/{\sf mass}(B_j)$ if $i \in B_j$ or $0$ otherwise, and (2) let
$y'_i = \min(1, (1+1/\alpha)y_i)$.

\begin{lemma}
  \label{lemma:lp-rounding-filtered-cost}
  Given an optimal primal solution $(x, y)$, there is a primal
  feasible solution $(x',y')$ such that (1) $\sum_{i} x'_{ij} = 1$,
  (2) if $x'_{ij} > 0$, then $d(j, i) \leq (1+\alpha)\delta_j$, and
  (3) $\sum_{i} f_iy_i \leq (1 + \frac1\alpha)\sum_i f_iy'_i$.
\end{lemma}
\begin{proof}
  By construction, (1) clearly holds. Furthermore, we know that if
  $x'_{ij} > 0$, it must be the case that $i \in B_j$, so $d(j, i)
  \leq (1+\alpha)\delta_j$, proving (2). By definition of $y'_i$,
  $\sum_{i} f_iy_i \leq (1 + \frac1\alpha)\sum_i f_iy'_i$, proving
  (3). Finally, since in an optimal LP solution, $\sum_{i} x_{ij} =
  1$, we know that ${\sf mass}(B_j) \geq \frac{\alpha}{1 + \alpha}$,
  by an averaging argument. Therefore, $x'_{ij} \leq (1 +
  \frac1\alpha)x_{ij} \leq \min(1, (1 + \frac1\alpha)y_i) = y'_i$,
  showing that $(x', y')$ is primal feasible.
\end{proof}

\smallskip
\noindent\textbf{Rounding:} The rounding phase is more challenging
to parallelize because it is inherently sequential---a greedy
algorithm which considers the clients in an increasing order of
$\delta_j$ and appears to need $\Omega(n_c)$ steps.  We show, however,
that we can achieve parallelism by eagerly processing the clients $S =
\{j : \delta_j \leq (1+\vareps)\tau\}$.  This is followed by a
clean-up step, which uses the dominator-set algorithm to rectify the
excess facilities.  We precompute the following information: (1) for
each $j$, let $i_j$ be the least costly facility in $B_j$, and (2)
construct $H = (C, F, ij \in E \text{ iff. } i \in B_j)$.

There is a preprocessing step to ensure that the number of rounds is
polylogarithmic in $m$. Let $\theta$ be the value of the optimal LP
solution. By an argument similar to that of
Section~\ref{sec:fac-loc-greedy}, we can afford to process all clients
with $\delta_j \leq \theta/m^2$ in the first round, increasing the
final cost by at most $\theta/m \leq \opt/m$.  The algorithm then
proceeds in rounds, each performing the following steps:
\begin{algo}[h]
  \begin{enumerate}
    \denselist

  \item Let $\tau = \min_j \delta_j$.

  \item Let $S = \{j : \delta_j \leq (1+\vareps)\tau\}$ and 

  \item Let $J = \udom(H)$, add $I = \{i_j : j \in J\}$ to $F_A$;
    finally, remove all of $S$ and $\cup_{j \in S} B_j$ from $V(H)$.
  \end{enumerate}
  \vspace{-2mm}
\end{algo}

Since $J$ is $U$-dominator of $H$, we know that for all distinct $j,
j' \in J$, $B_j \cap B_{j'} = \emptyset$; therefore, $\sum_{i \in I}
f_i = \sum_{j \in J} f_{i_j} \leq \sum_{j \in J} \big(\sum_{i \in B_j}
x'_{ij}f_{i_j}\big) \leq \sum_{j \in J} y'_if_{i_j} \leq \sum_{j \in J}
y'_if_{i}$, proving the following claim:
\begin{claim}
  \label{claim:lp-rounding-fac-cost}
  In each round, $ \sum_{i \in I} f_i \;\; \leq \;\; \sum_{i \in
    \cup_{j} B_j} y'_if_i$.
\end{claim}

Like our previous analyses, we will define a client-to-facility
assignment $\pi$ convenient for the proof. For each $j \in C$, if $i_j
\in F_A$, let $\pi_j = i_j$; otherwise, set $\pi_j = i_{j'}$, where
$j'$ is the client that causes $i_j$ to be shut down (i.e., either
$i_j \in B_{j'}$ and $j'$ was process in a previous iteration, or both
$j$ and $j'$ are processed in the same iteration but there exists $i
\in B_j \cap B_{j'}$).

\begin{claim}
  \label{claim:lp-rounding-connection-cost}
  Let $j$ be a client. If $i_j \in F_A$, then $d(j, \pi_j) \leq
  (1+\alpha)\delta_j$; otherwise, $d(j, \pi_j) \leq
  3(1+\alpha)(1+\vareps)\delta_j$.
\end{claim}
\begin{proof}
  If $i_j \in F_A$, then by
  Lemma~\ref{lemma:lp-rounding-filtered-cost}, $d(j, \pi_j) \leq
  (1+\alpha)\delta_j$. If $i_j \not\in F_A$, we know that there must
  exist $i \in B_j$ and $j'$ such that $i \in B_{j'}$ and $\delta_{j'}
  \leq (1 + \vareps) \delta_j$. Thus, applying
  Lemma~\ref{lemma:lp-rounding-filtered-cost} and the triangle
  inequality, we have $d(j, \pi_j) \leq d(j,i) + d(i, j') + d(j',
  i_{j'}) \leq 3(1+\alpha)(1+\vareps)\delta_j$.
\end{proof}

\smallskip
\noindent\textbf{Running Time Analysis:} The above algorithm will terminate in
at most $O(\log_{1+\vareps} m)$ rounds because the preprocessing step
ensures the ratio between the maximum and the minimum $\delta_j$
values are polynomially bounded. Like previous analyses, steps 1 -- 2
can be accomplished in O(1) basic matrix operations, and step 3 in
$O(\log m)$ basic matrix operations on matrices of size $m$.  This
yields a total of $O(\log_{1+\vareps} m \log m)$ basic matrix
operations, proving the following theorem:

\begin{theorem}
  Given an optimal LP solution for the primal LP in
  Figure~\ref{fig:primal-dual-program}, there is an $\RNC$ rounding
  algorithm yielding a $(4+\vareps)$-approximation with $O(m
  \log m \log_{1+\vareps} m)$ work.  It is cache efficient.
\end{theorem}




\section{\texorpdfstring{\lowercase{{\large $k$}}-Median}{k-Median}: Local Search}

\label{sec:k-median}

Local search, LP rounding, and Lagrangian relaxation are among the
main techniques for approximation algorithms for $k$-median.  In this
section, building on the algorithms from previous sections, we present
an algorithm for the $k$-median problem, based on local-search
techniques.  The natural local-search algorithm for $k$-median is very
simple: starting with any set $F_A$ of $k$ facilities, find some $i
\in F_A$ and $i' \in F \setminus F_A$ such that swapping them
decreases the $k$-median cost, and repeat until no such moves can be
found.  Finding an improving swap or identifying that none exists
takes $O(k(n-k)n)$ time sequentially, where $n$ is the number of nodes
in the instance.  This algorithm is known to be a
$5$-approximation~\cite{AryaGKMMP:siamjc04,GuptaT:arxiv08}.

The key ideas in this section are that we can find a good initial
solution $S_0$ quickly and perform each local-search step
fast. Together, this means that only a small number of local-search
steps is needed, and each step can be performed fast.  To find a good
initial solution, we observe that any optimal $k$-center solution is
an $n$-approximation for $k$-median.  Therefore, we will use the
$2$-approximation from Section~\ref{sec:k-center} as a factor-$(2n)$
solution for the $k$-median problem.  At the beginning of the
algorithm, for each $j \in V$, we order the facilities by their
distance from $j$, taking $O(n^2\log n)$ work and $O(\log n)$ depth.

Let $0 < \vareps < 1 $ be fixed. We say that a swap $(i, i')$ such
that $i \in F_A$ and $i' \in F\setminus F_A$ is \emph{improving} if
$\kmed(F_S - i + i') < (1-\beta/k)\kmed(F_S)$, where $\beta =
\vareps/(1+\vareps)$.  The parallel algorithm proceeds as follows.  In
each round, find and apply an improving swap as long as there is
one. We now describe how to perform each local-search step
fast. During the execution, the algorithm keeps track of $\varphi_j$,
the facility client $j$ is assigned to, for all $j \in V$.  We will
consider all possible test swaps $i \in F_A$ and $i' \in V\setminus
F_A$ \emph{simultaneously in parallel}.  For each potential swap $(i,
i')$, every client can independently compute $\Delta_j = d(j, F_A - i
+ i') - d(j, F_A)$; this computation trivially takes $O(n_c)$ work and
$O(1)$ depth, since we know $\varphi_j$ and the distances are
presorted.  From here, we know that $\kmed(F_A - i + i') - \kmed(F_A)
= \sum_j \Delta_j$, which can be computed in $O(n)$ work and $O(\log
n)$ depth.  Therefore, in $O(k(n-k)n)$ work and $O(\log n)$ depth, we
can find an improving swap or detect that none exists. Finally, a
round concludes by applying an improving swap to $F_A$ and updating
the $\varphi_j$ values.

Arya et al.~\cite{AryaGKMMP:siamjc04} show that the number of rounds
is bounded by \[ O\left(\log_{1/(1-\beta/k)}
  \big({\kmed(S_0)}/{\opt}\big)\right) = O\left(\log_{1/(1-\beta/k)}
  (n)\right)
\] Since for $0 < \vareps <1$, $ \ln\big({1}/(1 - \beta/k)\big) \leq
\frac2k\ln \left(1/(1 - \beta)\right)$, we have the following theorem,
assuming $k \in O({\sf polylog}(n))$, which is often the case in many
applications:
\begin{theorem}
  For $k \in O({\sf polylog}(n))$, there is an \NC
  $O(k^2(n-k)n\log_{1+\vareps} (n))$-work algorithm which gives a
  factor-$(5 + \vareps)$ approximation for $k$-median.
\end{theorem}


\noindent\emph{Remarks.} Relative to the sequential algorithm, this algorithm
is work efficient---regardless of the range of $k$.  In addition to
$k$-median, this approach is applicable to $k$-means, yielding an
$(81+\vareps)$-approximation~\cite{GuptaT:arxiv08} in general metric
spaces and a $(25+\vareps)$-approximation for the Euclidean
space~\cite{KMNPSW03}, and the same parallelization techniques can be
used to achieve the same running time.  Furthermore, there is a
factor-$3$ approximation local-search algorithm for facility location,
in which a similar idea can be used to perform each local-search step
efficiently; however, we do not know how to bound the number of
rounds.

\section{Conclusion}
\label{sec:concl}

This paper studies the design and analysis of parallel approximation
algorithms for facility-location problems, including facility
location, $k$-center, $k$-median, and $k$-means.  We presented several
efficient algorithms, based on a diverse set of approximation
algorithms techniques. The practicality of these algorithms is a
matter pending experimental investigation.


 \small 
  \bibliographystyle{alpha}
  \bibliography{../ref,../abbrev,../embedding}

\end{document}